\documentclass[12pt]{article}
\usepackage{epsfig}
\begin{document}
\title{Cross-over from reptation to Rouse dynamics in a 1-dimensional model}
\author{Andrzej Drzewi\'nski \\
Czestochowa University of Technology, \\
Institute of Mathematics and Computer Science,\\
ul.Dabrowskiego 73, 42-200 Czestochowa, Poland\\*[4mm]
J.M.J. van Leeuwen \\
Instituut-Lorentz, University of Leiden, P.O.Box 9506, \\
2300 RA Leiden, the Netherlands} 
\maketitle

\begin{abstract}
A simple 1-dimensional model is constructed for polymer motion. It exhibits the cross-over 
from reptation to Rouse dynamics, through gradually allowing hernia creation and annihilation.
The model is treated by the density matrix technique which permits an accurate 
finite-size-scaling analysis of the behavior of long polymers.
\end{abstract} 

\section{Introduction}

It is known for some time that the universal properties of long polymers can be obtained
from stochastic lattice models, which in themselves are rather crude representations of the 
intricate polymeric motion. The reason is that long polymers are critical \cite{deGennes}
and in critical systems the universal properties are independent of the microscopic details.
There are basically two modes of motion for polymers. One is reptation, which is the 
mechanism for polymers dissolved in a gel and to a lesser extent for dense polymer melts.
Here the polymer is strongly confined and the main degree of freedom is motion inside the
confining tube. The other mode applies for dilute solutions where the polymers can also move
freely sideways. This is usually called Rouse dynamics. One can easily envision situations
where a mix of these two mechanisms is present.

The interesting aspect is that the two modes have different and unusual dynamical exponents.
In polymer motion the dynamic exponent is related to the renewal time $\tau$. It increases as
$\tau \sim N^z$, where $z$ is the dynamic exponent and $N$ a measure for the length of the
chain. Whereas many models have the dynamic exponent $z=1$, showing isotropy 
between time and space, reptation has an exponent $z=3$.  This value has been a bit 
controversial, since viscosity measurements point at $z=3.4$, whereas the theories agree 
on $z=3$. The  discrepancy has recently been removed \cite{Carlon1} by an accurate 
finite-size-scaling analysis using the density-matrix technique (DMRG) introduced by 
White \cite{White1}. 
Rouse dynamics on the other hand is found to have the exponent $z=2$. It is therefore 
interesting to study the cross-over between these two mechanisms and to find out how a 
mixing-in of Rouse dynamics changes the dynamical exponent from $z=3$ to $z=2$. 

The most convenient model for reptation is the Rubinstein-Duke (RD) model designed by 
Rubinstein \cite{Rubinstein} and extended by Duke \cite{Duke}, by introducing a driving field. 
The mobile units, the reptons, only move along the tube that the chain traces out in the lattice.  
The advantage of the model is that the dimension $d$ of the lattice, in which the polymer chain
is embedded, becomes a parameter, which influences the behavior of the ends of the chain
but not of the bulk. As this parameter $d$ is one of the details, having no 
influence on the universal properties, one often studies \cite{Widom} the case $d=1$, although 
the RD model becomes somewhat artificial in a 1-dimensional embedding. 

There are two forms of sideways motion of the reptons. When a cell is occupied by three reptons
the middle one can enter a neighboring cell without crossing a barrier. This is called hernia 
creation and the opposite process is hernia annihilation. The other forms of sideway motion 
imply that the chain crosses a barrier. These are the typical motions allowed in Rouse dynamics.
Within the spirit of the physics of the RD model, hernia 
creation and annihilation should be allowed, but that makes the model essentially more 
difficult. For instance, the role of the embedding dimension cannot be simply reduced to a 
parameter $d$, influencing only the ends of the chain. 

The usual argument to omit the hernia creation and annihilation, is that these processes do
not alter the universal properties. This is likely to be true in larger $d$, where hernias 
become a fraction of the possibilities for the chain, but in $d=1$ they are of major importance
as we will show in this paper. In fact the hernia creation and annihilation mimic the role
of Rouse dynamics in a 1-dimensional embedding and therefore it is a convenient mechanism 
to study the cross-over from reptation to Rouse dynamics.

The dynamic exponent $z$ is obtained from the gap in the spectrum of the Master Equation.
Apart from this gap another interesting quantity is the diffusion coefficient. We obtain this
from the  model by studying the drift velocity in the limit of a weak driving field. For 
reptation the diffusion coefficient decays as $N^{-2}$ for chains of length $N$, while for 
Rouse dynamics the diffusion is speeded up to $N^{-1}$. Next to the cross-over of the 
dynamic exponent, we study in this paper the cross-over of the diffusion exponent.

\section{The model}

The model is a 1-dimensional chain of $N+1$ reptons, connected by $N$ links, 
$(y_1, \cdots, y_N)$. The links are either in the forward direction, $y_i = 1$, or in the
backward direction $y_i = -1$, or have the value $y_i=0$. The cases $y_i = \pm 1$ are 
considered as taut links, while $y_i = 0$ is a slack link or an element of stored length. 
The basic motion rule is the hopping of this stored length unit along the chain, 
by interchanging with taut links. If it moves in the forward
direction, its transition rate is biased by a factor $B>1$, while the hopping rate in the 
backward direction is decreased by the factor $B^{-1} <1$. The biases represent an 
external field driving the reptons of the chain. At the end of the chain the
links may change from  slack to taut and vice versa, thereby adding or subtracting
an element of stored length, again with a bias depending on the direction of the transport
of length. These motion rules form the much studied Rubinstein-Duke (RD) model. Our 
new element is that we allow a neighboring pair of opposite taut links to change 
into a pair of slack links and vice versa. We describe this as the annihilation viz.
creation of a hernia. The transition rate for hernia creation/annihilation is $h$,
multiplied with a bias based on the sign of motion of the middle repton  of the hernia. 

Without hernia motion the RD model is a typical model for reptation. The tube, which is 
the sequence of taut links, can only be changed from the ends. This is a slow 
process, since the taut links in the bulk have to wait till they happen to drift to one of the 
ends, before they can change their value. Simple counting tells that the 
inner taut links need at least $N^2$ repton moves, if they could renew themselves 
in a systematic way. The change of 
configuration is however a diffusive process in configuration space and therefore the
average renewal time is $N^4$ measured in single repton moves, or $N^3$ in chain 
updates. So the reptation renewal time $\tau \sim N^3$. Obviously hernia creation and 
annihilation speed up the renewal of the chain and the point of this note is to see how 
they can overtake the reptation mechanism.

A similar global argument \cite{Widom} yields that the pure RD model (without hernia 
creation/annihilation) has a drift velocity decaying as $N^{-1}$, leading to an asymptotic 
$N^{-2}$ behavior for the diffusion coefficient. 

\section{The Master Equation}

Our model is, as all the hopping models, governed by the Master Equation for the 
probability distribution $P ({\bf Y})$ where ${\bf Y}$ stands for the complete 
configuration $(y_1, \cdots, y_N)$. It has the form
\begin{equation} \label{a1}
{\partial P({\bf Y},t) \over \partial t} =  \sum_{\bf Y'}\left[ W ({\bf Y} | {\bf Y}')
P ({\bf Y}',t) - W ({\bf Y}' | {\bf Y}) P({\bf Y},t)\right]
\equiv \sum_{\bf Y'} M({\bf Y},{\bf Y}') P({\bf Y}',t).
\end{equation}
The $W$'s are the transitions rates and  
the matrix $M$ contains the gain terms (in the off-diagonal elements) and the loss
terms (on the diagonal). Conservation of probability implies that the sum over the
columns of the matrix vanishes. So the matrix has a zero eigenvalue and the 
eigenfunction corresponding to this eigenvalue is the stationary state of the system,
to which every other initial state ultimately decays. The matrix is non-symmetric,
due to the bias, which gives different rates to a process and its inverse. Thus one has 
to distinguish between left and right eigenfunctions. The left eigenfunction belonging
to the zero eigenvalue is trivial (all components equal); the problem is to find the 
right eigenfunction as the stationary state probability distribution. 

The renewal time is given by the slowest decaying eigenstate. Thus the gap in the
spectrum near 0 is the inverse renewal time. All eigenvalues must have of course
a negative real part, otherwise probability would grow unlimited. The form (\ref{a1})
stresses the similarity to quantum mechanical problems. Indeed the linear structure
of the polymer chain makes it a 1-dimensional quantum problem, however, with a 
non-hermitian hamiltonian. Our approach to the solution exploits this analogy by 
applying the so-called DMRG method for quantum problems to find the properties
of the transition matrix $M$. In previous publications \cite{Carlon2} the application of 
this method to polymer motion has been described in detail. Here we present only the results.

We confine ourselves to the renewal time and the diffusion coefficient. The renewal time
is usually defined at zero driving field. Also the standard diffusion coefficient refers to 
zero driving field . However, to determine the diffusion coefficient we must turn on an
infinitesimal field and compute the drift velocity. This can be done by expanding 
the Master Equation with respect to the field. The field enters in the bias $B$ which we
represent as
\begin{equation} \label{a2}
B = \exp (\epsilon/2),
\end{equation} 
where $\epsilon$ is a dimensionless parameter measuring the field strength. Then we
expand the Master Equation in powers of $\epsilon$
\begin{equation} \label{a3}
{\cal M} = {\cal M}_0 + \epsilon {\cal M}_1 + \cdots , \quad \quad \quad 
P ({\bf Y}) = P_0 ({\bf Y}) + \epsilon P_1 ({\bf Y}) + \cdots 
\end{equation} 
and  obtain the equations
\begin{equation} \label{a4}
{\cal M}_0 P_0 = 0 , \quad \quad \quad 
{\cal M}_0 P_1 = - {\cal M}_1 P_0.
\end{equation} 
The first equation is trivially fulfilled by a constant $P_0 ({\bf Y})$, since the matrix 
$M_0$ is symmetric and the right eigenvector becomes equal to the trivial left eigenvector.
The second equation is a set of homogeneous linear equations for the components of
$P_1 ({\bf Y})$. It is soluble, since the right hand side of the equation is perpendicular 
to the left eigenvalue (which remains true to all orders is $\epsilon$). So we can make the
solution definite by requiring that it is also orthogonal to the trivial left eigenvector.
$P_1 ({\bf Y})$ yields the lowest order drift velocity $v_d$ and the diffusion constant
follows by the Einstein relation as
\begin{equation} \label{a5}
D = {1 \over N} \,\left( {\partial v_d \over \partial \epsilon }\right)_{\epsilon=0}. 
\end{equation} 
\begin{figure}[h]
\begin{center}
    \epsfxsize=12cm%\linewidth
    \epsffile{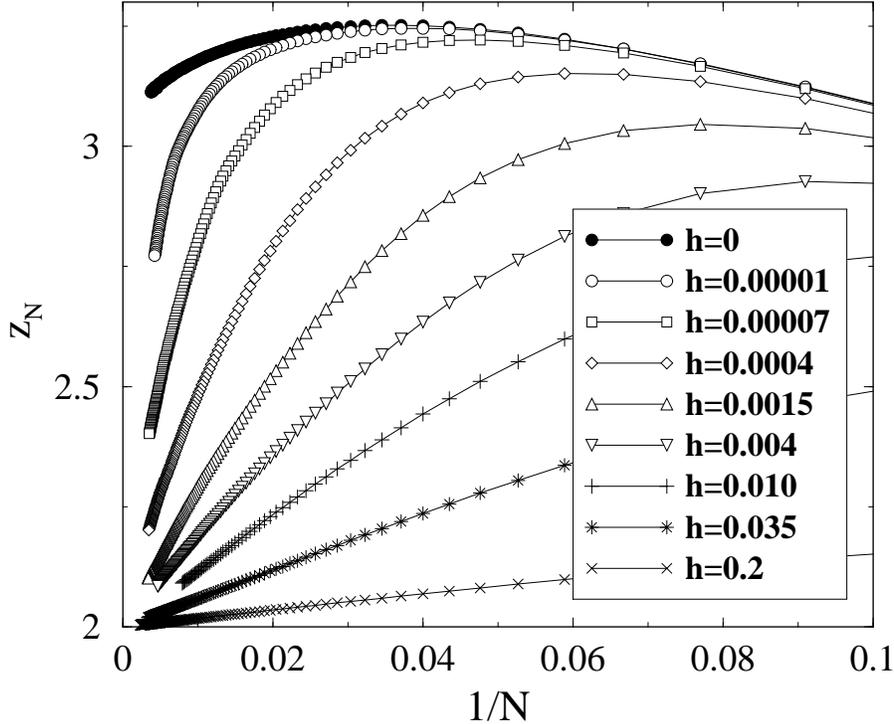}
    \caption{The renewal time as function of the length of the chain for various values of the
hernia creation/annihilation rate $h$.}  
\label{renewal}
\end{center}
\end{figure} 

\section{Scaling exponents}

\begin{figure}[h]
\begin{center}
    \epsfxsize=12cm%\linewidth
    \epsffile{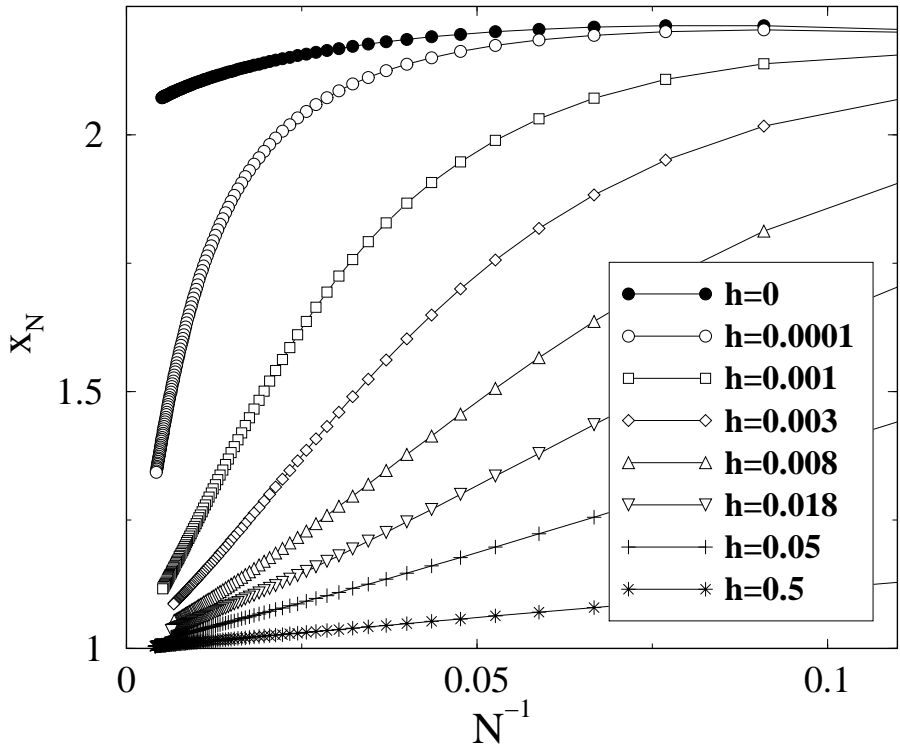}
    \caption{The diffusion exponent $x_N$  as function of the length of the chain for various 
values of the hernia creation/annihilation rate $h$.}  
\label{diffusion}
\end{center}
\end{figure}
One of the advantages of the DMRG method is that it calculates the properties, e.g.
the gap, for a growing length $N$ of the chain. In principle the method allows to go to any 
length, but it is in practice limited by instabilities and computational time.  
We have speeded up the process by using the field inversion symmetry in linear order 
in $\epsilon$, both for the gap and the diffusion coefficient. 
This make the results very well suited
for a finite-size-scaling analysis. We convert the gap as function of $N$ to a renewal time 
$\tau (N)$.   In Fig. \ref{renewal} we present the local exponent $z_N$, defined as 
\begin{equation} \label{b1}
z_N = {\ln \tau(N+1) - \ln \tau(N-1) \over \ln (N+1) - \ln (N-1)} 
\simeq {d \ln \tau \over d \ln N}.
\end{equation} 

The DMRG method gives values, accurate enough, such that the small differences in (\ref{b1})
do not spoil the accuracy. The various curves correspond to different values of $h$. 
Previously we found that it is most suggestive to plot $\tau$ as function of $N^{-1/2}$.
This applies indeed for the case $h=0$, but for non-zero 
values of $h$, a plot agains $N^{-1}$ gives more straight curves. Some features are noteworthy:
\begin{itemize}
\item Chains of the order of $N \simeq 100$ are not yet in the asymptotic 
regime \cite{Carlon2}. So there are large corrections to scaling.  
This is the origin of the earlier mentioned controversy between theory and 
experiment. In particular the plateau in the $h=0$ curve (the pure reptation case)
may easily lead to the conclusion that the exponent has settled on the (too large)
value. 
\item The influence of small values of $h$ is quite strong for long chains in particular 
for small values of $h$. We come back on this point when we discuss the cross-over 
behavior.
\item The asymptotic behavior of the exponent (for $N \rightarrow \infty$) differs
for $h=0$ from all the other curves. While the theoretical value $z_\infty = 3$ 
for reptation, is quite compatible with the data, it is definitely excluded for the curves
$h \neq 0$. They clearly point to the common value $z_\infty = 2$, which is characteristic
for Rouse dynamics.
\end{itemize}
In Fig. \ref{diffusion}  we plot in the same way the local exponent $x_N$ for the diffusion 
coefficient, defined as
\begin{equation} \label{b2}
x_N = -{\ln D(N+1) - \ln D(N-1) \over \ln (N+1) - \ln (N-1)} 
\simeq -{d \ln D \over d \ln N}.
\end{equation} 
The picture has a similar message as the previous one.
It is clear that, without hernia motion ($h=0$), the exponent evolves towards the
value $2$, while for any non-zero value of $h$, it aims at the value 1. Again one 
has large corrections to scaling. These corrections make it impossible to determine the 
exponent from ln-ln plots. Only due to the high accuracy of the DMRG method one
can derive exponents from formulae like (\ref{b1}) or (\ref{b2}).

\begin{figure}[h]
\begin{center}
    \epsfxsize=12cm%\linewidth
    \epsffile{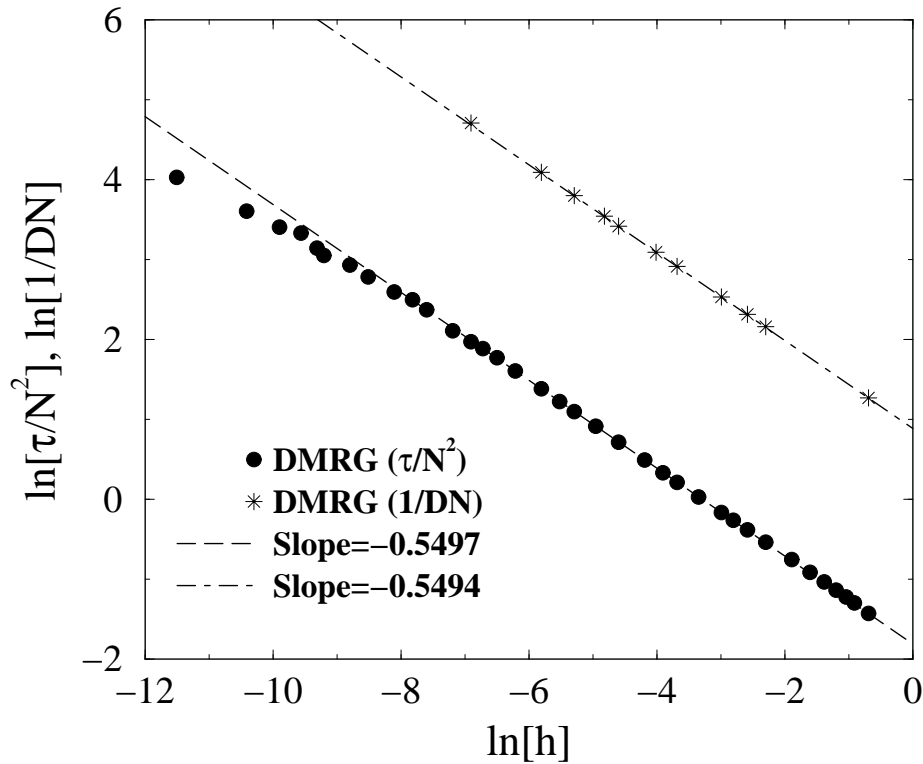}
    \caption{ln-ln plots of the renewal time and the diffusion coefficient as function of $h$}  \label{lnln}
\end{center}
\end{figure}

In Fig. \ref{lnln} we have made a plot of $\ln(\tau/N^2)$ and $-\ln(DN)$. 
In both cases the asymptotic values  $ N \rightarrow \infty$ are plotted as function of 
$\ln h$. As one sees the curves are fairly straight, with a slope -0.55, in the domain where 
the data are most accurate. For very small values of $h$ we see in the renewal data a somewhat 
smaller slope, a trend which is also detectable in the diffusion data on closer inspection. 
\section{Crossover scaling}

The point of crossover scaling is to represent the data for various values of $h$ in
one single curve. Anticipating the asymptotic values of the two regimes: 
$h \rightarrow 0$ and a fixed $h \neq 0$, the following representation is adequate
for the renewal time.
\begin{equation} \label{c1}
\tau (N, h) = N^3 g (h^\theta N).
\end{equation} 
The connection with the previous representation runs via the relation
\begin{equation} \label{c2}
{d \ln \tau \over d \ln N} = 3 + {d \ln g (h^\theta N) \over d \ln (h^\theta N)}.
\end{equation} 
We expect the cross-over function $g(x)$ to be expandable for small arguments as
\begin{equation} \label{c3}
g(x) = g_0 + g_1 x + \cdots
\end{equation} 
and for large arguments as
\begin{equation} \label{c4}
g(x) \simeq {1 \over x} \left( g_{-1} + {g_{-2} \over x} + \cdots \right).
\end{equation}
Inserting the asymptotic behavior (\ref{c4}) into (\ref{c1}) we obtain
\begin{equation} \label{c5}
\ln(\tau/N^2) = \ln g_{-1} - \theta \ln h + \cdots,
\end{equation} 
where the dots refer to corrections of order $1/N$. So the slope in Fig. \ref{lnln} gives the
value of $\theta$. 
The value $g(0)$ can be derived from a plot of $\tau N^{-3}$ versus $N^{-1}$. 
We find the value $g(0) \simeq  0.2$. 
\begin{figure}[h]
\begin{center}
    \epsfxsize=12cm%\linewidth
    \epsffile{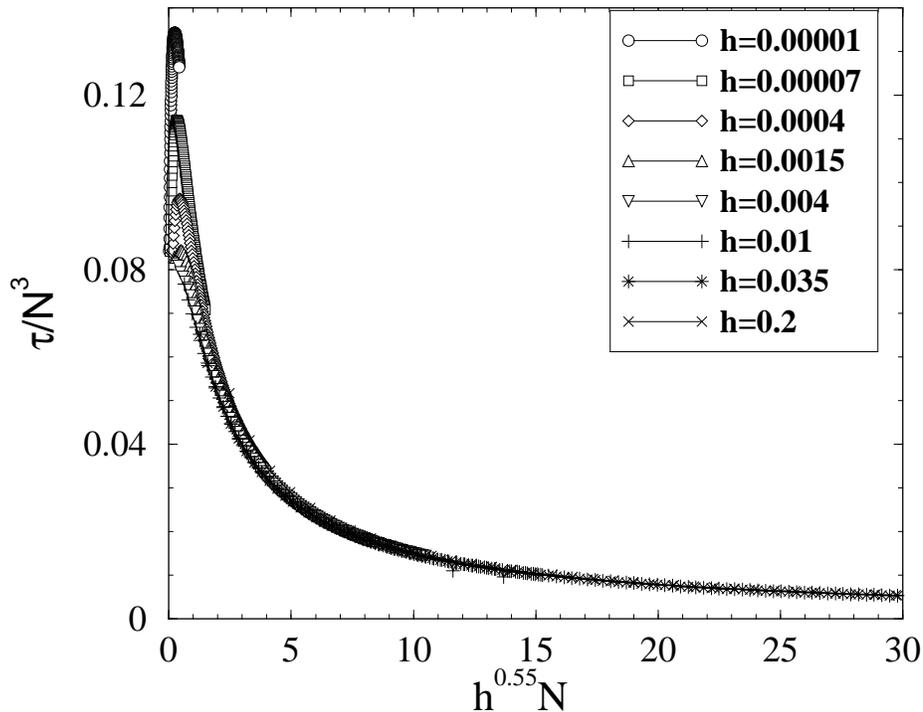}
    \caption{The cross-over function $g$ as defined  in (\ref{c1}) as function of the argument 
$h^{0.55} N$.}  
\label{crossren}
\end{center}
\end{figure}
In Fig. \ref{crossren} we have plotted the scaling
function $g$ as a function of $h^{0.55} N$. The observed data collapse is the proof for 
cross-over scaling. The deviation for small argument in Fig. \ref{crossren} are due to short 
chains. The scaling curve should aim, for small arguments,  at the value $g(0) \simeq 0.2$,
which we deduced from the $h=0$ curve. 

\begin{figure}[h]
\begin{center}
    \epsfxsize=12cm%\linewidth
    \epsffile{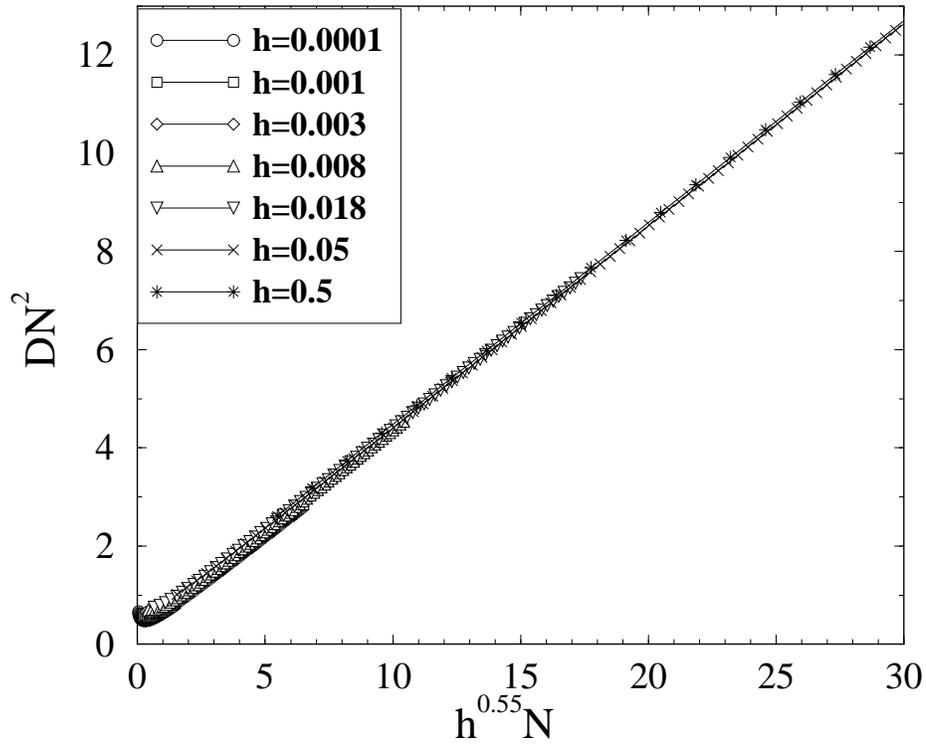}
    \caption{The cross-over function $f$ as defined  in (\ref{c2}) as function of the argument 
$h^{0.55} N$.}  
\label{crossdif}
\end{center}
\end{figure}

In Fig. \ref{crossdif} we plot similarly the diffusion coefficient in the form
\begin{equation} \label{c6}
D (N, h) = N^{-2} f(h^\theta N)
\end{equation}
with the same value $\theta =0.55$. As one sees the collapse is excellent. 
The cross-over scaling function $f$ approaches again a finite value at $x=0$. In view of
the data for the diffusion coefficient at $h=0$ we have $f(0) = 0.4$, which is quite consistent
with the behavior of the $h>0$ curves. For
large arguments, $f(x)$ should behave as $f(x \rightarrow \infty) \sim x$.

We hesitate to claim that the cross-over scaling exponent differs from the value $\theta = 1/2$,
which certainly gives a less perfect data collapse. An argument in favor of  $\theta = 1/2$ is 
based on the simple estimate of the times to remove a hernia for the two mechanisms. 
As we mentioned, pure reptation requires $N^4$ single repton moves to refresh the chain as a 
whole. On that timescale the hernias (in total of order $N$) in the chain are annihilated and 
replaced by others. So it takes $N^3$ repton moves to forget a hernia by reptation. 
On the other hand, direct change of 
a hernia by creation or annihilation goes with a rate $h/N$. The fastest process dominates and 
the competion is controlled by the ratio of the rates $(h/N)/(1/N^3) = hN^2$. So the 
crossover scaling function should be a function of the ratio $h N^2$. It might well be that the
real asymptotic value for $\theta = 1/2$ and that we see in the window, where we have data,
an effective exponent. This looks similar to the story of the renewal exponent itself, which 
also was estimated as 3.4, while the true theoretical value is 3. As we mentioned we see in Fig.
\ref{lnln} a tendency to a smaller slope for the very small $h$, which supports this 
possibility.

\section{Discussion}

We have presented a simple model which demonstrates the cross-over from reptation
to Rouse dynamics. In the Rubinstein-Duke model the links in the direction of the field 
and those against the field cannot interchange and this makes reptation a slow process.
In our 1-dimensional model, hernia annihilation and creation, allow the two types of 
links to interchange and therefore these obstacles can be overcome.
In that sense they play the same role as the tube changes which are typical for Rouse dynamics. 

In a paper by Sartoni and Van Leeuwen, \cite{Sartoni} the 1-dimensional reptation with hernia
creation and annihilation, has been connected to a simpler model of two types of particles,
which move independently of each other along the chain. They also conclude that
the diffusion coefficient decays as $N^{-1}$, but they have to stick to a hernia
creation and annihilation rate equal to the hopping rate of the reptons. 
Here we could vary this rate at will and therefore study the cross-over behavior.
In a forthcoming paper we have related their findings to the recently introduced necklace model 
\cite{Terranova, condmat}.

\end{document}